\documentclass[runningheads]{llncs}
\usepackage{graphicx}
\usepackage[utf8]{inputenc}
\usepackage{authblk}
\usepackage[numbers]{natbib}
\usepackage{graphicx}
\usepackage{mathdots}
\usepackage[english]{babel}
\usepackage[colorinlistoftodos]{todonotes}
\usepackage{amsmath}
\usepackage{amsfonts}
\usepackage{amssymb}
\usepackage{subfigure}
\usepackage{caption}
\usepackage{url}
\usepackage{algorithm}
\usepackage{algpseudocode}
\usepackage{comment} 
\usepackage{bm}

\begin{document}
%
\title{Adversarial Knapsack and Secondary Effects of Common Information for Cyber Operations}
%
\titlerunning{Adversarial Knapsack and Secondary Effects}
%

\author{ Jon Goohs\inst{1} 
 \and Georgel Savin\inst{2,3} \and 
  Lucas Starks\inst{2} \and Josiah Dykstra\inst{4} \and 
 William Casey\inst{2}
}

%
\authorrunning{ J. Goohs et al.}

%


\institute{ United States Navy
\and
United States Naval Academy\\
\email{ wcasey@usna.edu}
\and
Forțele Navale Române
\and
Trail of Bits
}

\def\flipit{{\sc Flip-It} }
\def\flipitns{{\sc Flip-It}}
\def\colblotto{{\sc Colonel Blotto} }
\def\wknap{{\it Weighted Knapsack} }
\def\aknap{{\it Adversarial Knapsack} }
\def\dknap{{\it Dueling Knapsack} }

\maketitle              
\begin{abstract}
Variations of the \flipit game have been applied to model network cyber operations.  While \flipit can accurately express uncertainty and loss of control, it imposes no essential resource constraints for operations.  {\it Capture the flag} (CTF) style competitive games, such as \flipit, entail uncertainties and loss of control, but also impose realistic constraints on resource use.  As such, they bear a closer resemblance to actual cyber operations. 
We formalize a dynamical network control game for CTF competitions and detail the static game for each time step. 
The static game can be reformulated as instances of a novel optimization problem called \aknap (AK) or \dknap (DK) when there are only two players.  We define the \aknap optimization problems as a system of interacting \wknap problems, and illustrate its applications to general scenarios involving multiple agents with conflicting optimization goals, e.g., cyber operations and CTF games in particular.
Common awareness of the scenario, rewards, and costs will set the stage for a non-cooperative game.  
Critically, rational players may second guess  that their AK solution---with a better response and higher reward---is possible if opponents predictably play their AK optimal solutions.   
Thus, secondary reasoning which such as belief modeling of opponents play
can be anticipated for rational players and will introduce a type of non-stability where players maneuver for slight reward differentials.
To analyze this, we provide the best-response algorithms and simulation software to consider how rational agents may heuristically search for maneuvers. We further summarize insights offered by the game model by predicting that metrics such as Common Vulnerability Scoring System (CVSS) may intensify the secondary reasoning in cyber operations.



\end{abstract}

\section{Introduction and Motivation}

Cyber attacks are characterized by unwilling and often unknown loss of control of software, hardware, or services associated with networked data and devices. 
At the same time, defense and offense operations are subject to resource constraints. 
Given these constraints, attackers and defenders are required to plan and allocate resource usage toward their objectives while leveraging any informational asymmetries in their favor.  
Motivated to better understand capture the flag (CTF) style competition games where players defend their networks by patching vulnerable software and attack other networks by crafting exploits of vulnerabilities, we seek to design realistic game theory to better understand CTF and dynamic cyber operation scenarios in general. 
For players to achieve objectives in these scenarios, several types of information become important for planning:  {\it vulnerabilities, exploits}, and {\it patches}.  
We describe what these are in greater detail in section \S\ref{sec:DNCG}. For now, the basic attack involves aiming an {\it exploit} against a specific {\it vulnerability}.  
The outcome depends on the patch status of the vulnerability, if the vulnerable component was {\it unpatched} the attack succeeds, otherwise the exploit will fail (when the vulnerability was {\it patched}).  The damage inflicted by a successful attack can vary greatly, but depends on the vulnerability targeted.  
Additionally, automation is common, where defenders can perform many patches and attackers can perform many exploits with software tools.  

Following prior work~\cite{goohs2022reducing} that identifies the \wknap as an important tool for efficient resourcing for defense, we extend this reasoning to include the attacker as they are subject to similar constraints and would also seek efficient resourcing. But, then, the question becomes: {\it if each player is capable of solving the other's knapsack problem (using intelligence or common knowledge of its parameters), might a player deviate from equilibrium attempting to out maneuver the other's solution?}  We refer to this type of strategic adjustment or second guessing as {\it secondary effects}.
Here, we extend the game theory to explore these secondary effects, in particular if both player attempt to stay one step ahead of the others, what equilibria results?  

To understand this scenario we define a novel agent optimization problem called \aknap.  
We illustrate how \aknap is the essential game played at each time step within CTF or network control games, more generally.  Then, we resolve methodology to calculate the equilibrium pair for each \aknap instance.

The mathematical analysis offers three insights.  First, the underlying combinatorial optimization problem and its reward landscape will offer significant maneuvering options for players seeking reward differentials.
Secondly, common information such as metrics of vulnerability score and patch cost, for which the Common Vulnerability Scoring System (CVSS) is an example, intensify secondary reasoning.  And thirdly, the effects of secondary reasoning are limited by a cyclic algebraic structure arising in finite problems.



\section{Background and Related Work}
\label{sec:BG} 
Platforms such as Hack the Box (HTB)~\cite{htb} and King of the Hill~\cite{bock2018king} provide students with the opportunity to practice cyber defense and penetration testing within a safe sandbox environment.  
The general objective of these platforms are to provide learning challenges for defense (patch and protect) and offense (exploit and hack) within environments of virtual machines networked in secure private networks.  
These systems offer practical instances of real cyber security problems and mirror the essential challenges of cyber security.  
Services such as HTB also offer competitive games such as capture the flag (CTF) challenges where multiple players conduct both defensive and offensive cyber operations within virtual networks.  CTFs are used to test operational skills, build teamwork, and advance organizational cyber security posture~\cite{ctf} by developing skilled workforce of ethical hackers, pen testers, and network defenders.  
These platforms also feature well developed video game interfaces that attract a general audience, but strive to retain challenging and timely complex problems.
As such, they have been used for research~\cite{savin2023battle} and have emerged as recruitment systems for companies seeking a cybersecurity work force to assess candidate skill level.



While many types of game theory have considered network control, we will focus on two types closest to our presented model: \colblotto games, and \flipit game forms.  \colblotto games, described by mathematician \'Emile Borel in 1921, were popularized within post WWII operations research science when the game received its name~\cite{gross1950continuous}. 
These game theoretic ideas are common within network, cyber-physical system, and radio spectrum security ~\cite{gupta2014three,etesami2019dynamic, min2017defense, ferdowsi2017colonel, schwartz2014heterogeneous,labib2015colonel}.
Key to the \colblotto game is that players have resource constraints and have to solve a resource allocation problem to ready battle fields, depending on the other player's selected allocation, a sum is split among the players and distributed.
This notion of resource constraint is important to manage the defensive and offensive resources during cyber operations, for example in managing defense resources strategically~\cite{tirenin1999concept, mcmorrow2010science}, under denial of service attack~\cite{wang2016recent}, or specifically to neutralize exploitation via patch planning~\cite{cavusoglu2006economics, goohs2022reducing}.  

While \colblotto captures the realistic notion of resource constraints in cyber operations, it struggles to accurately model cyber operations in two separate ways.  
First, the notion of conquering a resource is complex.  Note that some software vulnerabilities can lead directly to full access while others lead to partial or no access.
Rules that account for the vulnerability and the access they provide can enable more realistic dynamics. 
Recently {\it vulnerability scoring systems}, such as Common Vulnerability and Exposures (CVE) and Exploit Prediction Scoring System (EPSS)~\cite{jacobs2023enhancing}, attempt to measure the notion of risk or damage ~\cite{munaiah2016vulnerability}. Our model incorporates rules and commonly known hazard scores to be more realistic. 
Secondly, \colblotto games typically describe complete information scenarios, but many cyber operations are performed in stealth mode, leading to uncertainties of machine states which can be resolved with costly verification actions.

The \flipit game developed by van Dijk et al.~\cite{van2013flipit} in 2013 addresses the important aspect of uncertain machine control state.  While it models operational costs that flip or audit control states, it does so without any reasonable constraint.  \flipit has become widely used to model network control problems~\cite{laszka2014flipthem, liu2021flipit}, but its lack of resource constraints impute limitations to its realism.  Recently, reinforcement learning for \flipit scenarios have been proposed~\cite{oakley2019mathsf, greige2020reinforcement}, but without resource constraints, an optimal solution may not result in an actionable plan.  We argue its important to optimize subject to resource constraints.  

Our model, like~\flipit, contains informational asymmetries for both attacker and defender, however, we incorporate realistic resource budgets of \colblotto for each time step.  
Other researchers have considered this very same combination, Zhang et al. ~\cite{zhang2014stealthy} blend information asymmetry with resource constraints to consider network control. however, they consider only one type of attack and allow attackers complete information concerning defender network.  Our model improves upon those.  
First, we model all known attack types via common vulnerability, patch and exploit databases.  Multiple attacks in cyber is an important characteristic and has been observed to yield evolutionary solutions~\cite{casey_epi2014}.  By accounting for many attack options, we formalize a more realistic set of strategic objectives and identify the associated combinatorial optimizations.
Secondly, we allow defender to maintain private information, such as patch status or even asymmetric information such as deceptive honeypots.  Our more realistic model, with multiple vulnerabilities, inherent resource constraints and informational asymmetries, is better suited to contribute to realistic studies of CTF games and dynamic network control problems.   


Various scoring systems for vulnerabilities have gained traction in network defense.  One example is the Common Vulnerability Scoring System (CVSS) that measures three types of hazards, {\it base}, {\it temporal}, and {\it environmental}, on a scale of 1-10~\cite{cvssinattackgraphs}.  These measures provide critical security information on a range of common vulnerability enumeration.  
These measures are intended to help improve defensive decision making and planning, for example allowing organizations to determine the need for specific patches or mitigating actions to prevent attacks.
However, information like this is also useful to attackers. 
One timeless problem is the uncertainty of how releasing a Common Vulnerability information will shift the attacker's posture, whether it may speed up attacker timeline driven by a need to `use it our loose it' (once a patch is released and widely distributed).  Alternatively, patch releases may shift attacker's focus to other 0-day vulnerabilities.  Still, other attackers can use CVE information to design exploits, which remain effective so long as there are unpatched vulnerabilities.  To our knowledge, this paper presents the first game theory equilibrium results that incorporate how commonly known hazard scores such as CVEs effect the decision making for both rational attacker and defender.   

The \wknap problem has a long history, first occurring in the late 19th century as a number partitioning problem, mathematician Tobias Dantzig\footnote{Father of George Datzig known for the simplex method solving linear programming problems.} gave the problem its current name.   The general \wknap decision problem is known to be NP-complete, when the weight parameter is bounded the problem has efficient solutions with dynamic, integer, or binary programming.  \wknap has been used within encryption systems, however abandoned as users tended to select keys that avoid the inherent hardness~\cite{merkle1978hiding, shamir1982polynomial, beier2004random} and allowing efficient knapsack solutions (for bounded weight) to crack the encryption.
\wknap problems surface in CTF games in relation to efficient resource use~\cite{goohs2022reducing}, similarly we use binary programming to solve instances of \wknap.  Here, we consider that the attackers also solves a similar resource use problem.  Since common vulnerability information is known to both attacker and defender, it will help both in planning, for the attacker this means effective attack strategies.  To consider this and the secondary effects of maneuvering we define a variation of the Knapsack Optimization Problem called \dknap (DK).  The problem aims to capture dual opposing goals for both attacker and defender agents.  Additionally, we design analysis tools as algorithms, the main tool can update belief by re-weight and then resolve a knapsack problem, and we indicate how this would be used to generate best response in the \aknap problem for any given, yet fixed opponent strategy.  We can then use those tools to calculate a sequence of strategies, corresponding to each player reasoning out the other player's best response to each re-weighted optimization problem.  The sequence of strategies is a natural framework for rational players to consider in fictitious play or via simulation.  We illustrate examples showing the sequence of best responses strategies will cycles and identify a cycle length bound and discuss the corresponding minimax strategies, game values and equilibrium for each zero-sum time step.  

Our main contributions are:
\begin{itemize}
    \item Improvement of model realism for network control games mixing the constraints of \colblotto and uncertainties of \flipit.  These improvements provide a sound foundation to understand cyber operations.
    \item Formalizing the \aknap problem as the essential optimization problem related to the game played during each time step.
    \item Analysis of the {\it secondary effects} of common knowledge.  For \aknap, we postulate that best response provides insights for how rational players may maneuver as secondary effects.  We provide an efficient algorithm that calculates best response by re-weighing the original Knapsack problem. 
    \item We determine that the second order effects of \aknap are bound to cycle and therefore limited, we provide a method to calculate reward variation, suggest strategy maneuvers, and determine mixed minimax equilibrium pairs and game value within the sequence of best responses.
\end{itemize}

\section{Formal Model for Network Control, Cyber Operations, and Games}
\label{sec:DNCG}

In this section, we expand the definition of vulnerabilities, patches, and exploits and explain how they relate in cyber operations. We then formalize the game theory for network control by defining time steps, networks, nodes, temporal node states, control states and graph, topology, and agent action spaces.  We then detail the state update procedures, define agent utility, resource constraints, and the constrained optimization problems.  

A {\it vulnerability} can be thought of as either a fault or weakness in design or operation that jeopardizes the desired security properties. 
Currently, several organizations maintain voluminous listings of vulnerabilities, examples include Common Vulnerability and Exposures (CVE) listing nearly 200K distinct vulnerabilities.   {\it Exploits} are repeatable scripts or codes that target a specific vulnerabilities to achieve some level of system compromise.  Currently, hackers use multiple exploit libraries (contain databases of exploit scripts), for example the Metasploit framework contains over 2,000 distinct exploit scripts.  These exploit scripts can be used in a variety of ways including:  offensive cyber operations, defensive penetration testing, CTF competitions,  develop cyber operational tools, or facilitate criminal hacking activities.   Finally, {\it patches} are any code or component modifications that will mitigate the hazards of a specific {\it vulnerability}, thereby once applied the patch prevent exploitation.  

{\bf Vulnerabilities, exploits and patches, relations and life-cycles:}  Usually, code vulnerabilities are first discovered by developers or security researchers as software bugs or fault conditions.  Once known, knowledgeable individuals can can develop exploits or patches which may or may not be contributed to common knowledge.  Generally, vulnerability information is widely or commonly shared via efforts such as CVE, and developers can then design patches which can be deployed to keep end users safe.  However, hackers can also use CVE information to design exploits.  
If all components are fully patched, no exploits will be effective.  If no components are patched, all exploits targeting those components are effective if/when they are used.  Once a component is patched, at least one exploit is no longer effective.  When components are phased out and replaced by newer components: any patches or exploits for the older component cease to be effective, and vulnerability discovery and exploit generation must move on to consider the currently deployed components.
{\bf Basic cost structure:}  We will generally assume the cost to develop an exploit or patch is much greater than (by one to two orders of magnitude) the cost of applying a patch.  Exploit development requires understanding of the vulnerable component,  developing patches may entail complex side effects for systems and be costly to design. Generally, patch deployment is simpler and lower in cost.



\subsection{Game State }
Let time be discretized as:  $T = \{ t_0 + k \delta : k \in \{0, 1, 2, \cdots \} \}$, with $\delta$ the largest appropriate time step required to model network states.  
Throughout, we will use $k$ as the time step index.
Let  ${\cal V}_k := V = \{ v_{1}, v_{2} \ldots v_{M}  \}$ be the set of network nodes at time $k$. 
We will use a fixed set of nodes\footnote{In most dynamic graph problems, the node set can be held constant, edges can have temporal dependence expressing node gain and loss to the network. }  for every time step $k$ and throughout we use $j$ as the node index.   
Let ${\cal E}_k \subset {\cal V} \times {\cal V}$ be the set of edges, $(v,w) \in {\cal E}_k$ if node $w$ is connected to node $v$ at time $k$.  The edge set captures the momentary network topology\footnote{The network graph provides all possible single-hop communications in the network, whether utilized or not. The topology is a set of local neighborhood of $v_j$ at time $k$ be defined as:
${\cal N}_{jk} = \{  w \in {\cal V } : ( v_j, w ) \in {\cal E}_k \}$ for each node $j \in \{1, \ldots, M \}$}.

Let $H = \{h_1, h_2, \ldots, h_N \}$ be a fixed set of vulnerabilities,
throughout we will use $i$ as the vulnerability index.  Let 
$$
\phi_{ijk}  = \begin{cases}
1 \text{ if vul } i \text{ is present and unpatched on node } j \text{ at time } k \\
0 \text{ otherwise }.
\end{cases}
$$
These zero-one indicator variables above determine the possible fronts where an attack maybe effective at time step $k$.  This information will be private to the defender, but otherwise act like fields in a \colblotto  game (\S\ref{sec:BG}).

{\bf Network Control Graph:}
Within the dynamic scenario, we consider a set of agents ${\cal A} = \{ a_1, a_2, \cdots a_L \}$, and control function $\Gamma : {\cal V \times T \times A} \rightarrow \{0, 1 
\}$ that indicates the state of control for each network node: $\Gamma_{jke} = 1$, if node $j$ is controlled by agent $a_e$ at time step $k$.  Throughout, we will use $e$ for the agent index.
The control states $\Gamma$, as a global function will not be known to any single agent, but only known to omniscient nature.  In CTF games, the control state is verified by an independent {\it Scorebot agent}~\cite{bock2018king}.
For a single machine $v_1$ and two players $\{a_1, a_2\}$, the function $k \rightarrow  \Gamma_{11k}$ can be viewed as the control state of a \flipit game.  Within our model, the true control state is unknown to players until they attempt a patch action, analogous to {\it audit} or {\it flip} operations in \flipitns. 
Similar to \flipit, player actions may only provide momentary certainty because the private actions of other players can change control states at any time.
For convenience we define:
$\Gamma_{ji}$ to be the owner of node $j$, that is $\Gamma_{ji} = \{ e : \Gamma_{jie} = 1 \} $

{\bf Network Topology and instantaneous attack surface:}
Network topology is an important control factor, and arise in security designs including layered security, VPNs, and Firewalls.  
To demonstrate the minimax solution to the static game, we defer the topological considerations for later and need only a notion of {\it boundary} or {\it reachable nodes}.  
The zero one variables $r_{jk}$ indicate if a machine $j$ is reachable at time step $k$ (by any player).

\subsection{Common Information} 
\label{sec:CI}
Vulnerability scoring systems are the de facto common knowledge of vulnerabilities.  We will assume a function $H_i$ that measures hazard of vulnerability $i$.  Additionally, we will assume a commonly known patch cost function $C_i( x ) = b_i + m_i x$ to reflect a tooling cost $b_i$ and unit scaling cost $m_i$ to fix $x$ instances of vulnerability $i$.  Also, we will assume a commonly known attack cost function $d_i( x) = B_i + M_i$ which likewise has tooling and scaling costs.  We assume the average exploit cost parameters to be one to two magnitudes greater than average patch costs, however, in our simulations of random cost structure we generate each as independent uniform values in a range.   

We summarize the state variable in the game with table~\ref{tab:t2} below, and then define decision spaces for players.
\begin{table}[]
\begin{center}
 \begin{tabular}{|r r|c|c||c|c|c|c|c|c|c||l|} 
 \hline 
 \multicolumn{12}{|c|}{Parameters and State at time step $k$} \\
 \hline
 \multicolumn{4}{|c||}{ rows are vuls } & \multicolumn{7}{|c||}{ columns are machines/nodes } & \\ 
 \hline 
\multicolumn{2}{|c|}{patch cost}  & haz & vul & $1$ & $2$ & $3$ & $\ldots$ & $j$ & $\ldots$ & $M$ & v tot \\
 \hline
 \hline
 $b_1$ & $m_1$ & $H_1$ & ${1}$ & ${\phi}_{11k}$ & $\phi_{12k}$ & $\phi_{13k}$ & $\ldots$ & ${\phi}_{1jk}$ & $\ldots$& $\phi_{1Mk}$ & $z_{1k}$ \\
 \hline
 $b_2$ & $m_2$ & $H_2$ &  $2$ & $\phi_{21k}$ & $\phi_{22k}$ & $\phi_{23k}$ & $\ldots$ & $\phi_{2jk}$ & $\ldots$& $\phi_{2Mk}$ & $z_{2k}$ \\
 \hline
 $b_3$ & $m_3$ & $H_3$ & $3$ & $\phi_{31k}$ & $\phi_{32k}$ & $\phi_{33k}$ & $\ldots$ & $\phi_{3jk}$ & $\ldots$& $\phi_{3Mk}$ & $z_{3k}$ \\
 \hline
 $\vdots$ & $\vdots$ & $\vdots$ & $\vdots$ & $\vdots$ &  $\vdots$& $\vdots$ & $\ddots$ &  $\vdots$ & $\iddots$ &  $\vdots$ & $\vdots$ \\
 \hline
 $b_i$ & $m_i$ & $H_i$ & $i$ & $\phi_{i2k}$ & $\phi_{i2k}$ & $\phi_{i3k}$ & $\ldots$ & $\phi_{ijk}$ & $\ldots$ & $\phi_{iMk}$ & $z_{ik}$ \\
 \hline
  $\vdots$ & $\vdots$ & $\vdots$ & $\vdots$ & $\vdots$ &  $\vdots$& $\vdots$ & $\iddots$ &  $\vdots$ & $\ddots$ &  $\vdots$ & $\vdots$ \\
 \hline $b_N$ & $m_N$ & $N$ & $h_N$ & $\phi_{N1k}$ & $\phi_{N2k}$ & $\phi_{N3k}$ & $\ldots$ & $\phi_{Njk} $ & $\ldots $ & $\phi_{NMk}$ & $z_{Nk}$ \\
 \hline
 \multicolumn{4}{|c|}{ node totals } & $\sigma_{1k}$ & $\sigma_{2k}$ & $\sigma_{3k}$ & $\ldots$ & $\sigma_{jk}$  & $\ldots $ & $\sigma_{Mk}$ & $Z_k$ \\
 \hline
 \multicolumn{4}{|c|}{ control states } & $\Gamma_{1k}$ & $\Gamma_{2k}$ & $\Gamma_{3k}$ & $\ldots$ & $\Gamma_{jk}$  & $\ldots $ & $\Gamma_{Mk }$ &  \\
   \hline
  \multicolumn{4}{|c|}{ accessible  } & $r_{1k}$ & $r_{2k}$ & $r_{3k}$ & $\ldots$ & $r_{jk}$  & $\ldots $ & $r_{Mk}$ &  \\ 
  \hline 
\end{tabular}
\end{center}
    \caption{Networks include static parameters for vulnerabilities, and variables $\phi_{ijk}$ (mutable by controlling agent) that indicate the presence of vulnerably $i$ on machine $j$ at time $k$.  Additionally, the control state for node $j$ at time $k$ is denoted $\Gamma_{jk}$ and will indicate the agent control for the node. The values $z_{ik}$ represents the aggregate hazard of vulnerabilities $i$ at time $k$.  the values $\rho_jk$ represents the aggregate hazard for node $j$ at time $k$.  These settings and topology, will determine the Colonel Blotto scenario occurring at each time point $k$.}
    \label{tab:t2}
\end{table}
In table~\ref{tab:t2}, Machines (or nodes) are represented by columns $1$ through $M$, and vulnerabilities are represented by rows $1$ through $N$.  
Vulnerabilities will have static characteristics, including a hazard score $H_i$ for vulnerability $i$, and a linear cost structure to patch instances: $c_i( x ) = b_i + m_ix$. 

For a fixed hazard function that assigns $H_i$ to vulnerability $i$, variables $\sigma_{jk} = \sum_{ i= 1 }^N \phi_{ijk} H_i$, and $z_{ik} = H_i \sum_{ j = 1}^M \phi_{ijk}$ will respectively represent the sum hazard of machine $j$ at time $k$, and the sum hazard presented by vulnerability $i$ at time $k$.   The value $Z_k = \sum_j \sigma_{jk} = \sum_i z_{ik}$ is the sum of all hazards due to vulnerabilities within the network present at time $k$.  In the next row, we denote the control states variables $\Gamma_{jk}$. 
The final row will indicate reachability of the node, or that it is within play reachable by attackers.  The control variables $\Gamma$ indicates the principal agent capable of muting column entries of $\phi$.
The defender (owner) $\Gamma_{jk}$ can for any vulnerability  mute $\phi_{ijk}$ from zero to one at time $k$ by patching the vulnerable component, this operation will cost the agent according to the cost structure of vulnerability $i$.   
Note we also formalize reversions, or patch removal, thereby reintroducing a fixed vulnerability, meaning the owner can mute an entry of $\phi$ from zero to one.  

\subsection{Player Roles and Action Space}
Players can perform {\it defense} actions such as patching or changing a vulnerability's patch status, {\it attack} actions such as launching an exploit.   A patch action removes other player's (attacker) ability to exploit the vulnerability, and thereby blocks the attacker's ability to flip the node if they select that vulnerability to exploit.  The patch status of any vulnerability on any node of the defender's network is private information, only known to the agent controlling the node (i.e., defender).  
We use the following indicator variables to indicate a defensive action:
$$
\zeta_{ijk} = 
\begin{cases}
1 \text{ if player changes patch status of vul } i \text{ on node } j \text{ at time } k \\ 
0 \text{ o.w. }
\end{cases}
$$
The action $\zeta_{ijk}$ can only be performed by $\Gamma_{jk} $.
Indicator variables $\zeta_{ijke}$ can be used when the player acting $e$ needs to be identified within the indicator function.   

In attack, we will assume each player can 
remotely probe or scan reachable nodes to determine software components 
but not its patch status.  
The player's action space will consist of selecting a machine and a vulnerability to deploy an exploit against.  
$$
\xi_{ijk} = 
\begin{cases}
1 \text{ if player attacks vul } i \text{ on node } j \text{ at time } k \\ 
0 \text{ o.w. }
\end{cases}
$$
Indicator variables $\xi_{ijke}$ can be used to further indicate the attacking player.

\begin{figure}[htb]
\centering
 \subfigure[Networks comprised of nodes, nodes comprised of components know to all players ]{ \includegraphics[height=40mm]{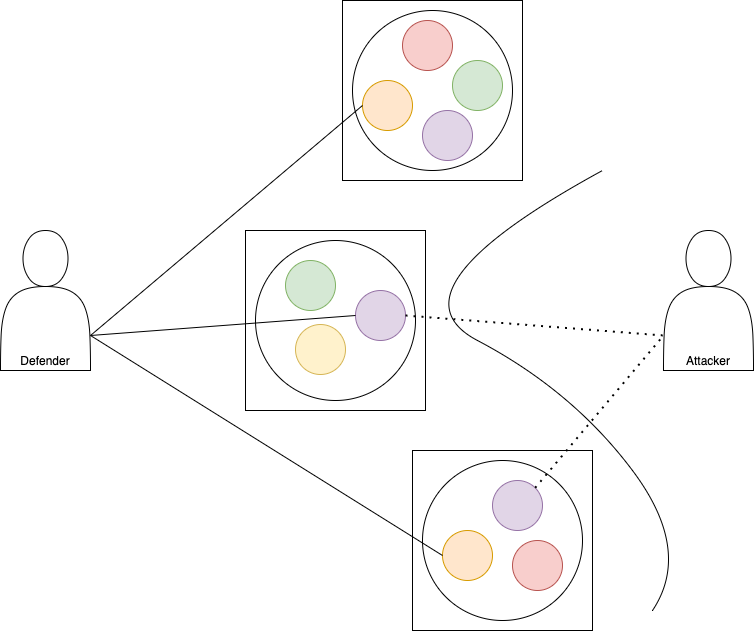}\label{fig:pwnit} } 
 \subfigure[Defender controls patch level privately, attack outcomes vary given patch level]{\includegraphics[height=40mm]{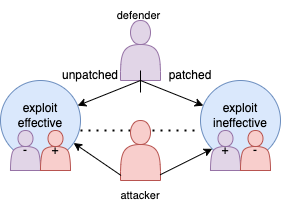}\label{fig:pwnit2}}  
  \caption{{\it Attack and Defense Asymmetries:} In Cyberspace, attackers have an advantage that they only need to exploit one vulnerability, while defenders must protect all vulnerabilities to remain secure.  
 (a) Networks are comprised of nodes. A node, owned by the defender, is a heterogeneous mix of software components, each component is visible to all players.  Components relate to fields in a \colblotto games, however the differ as well, as the attacker, only needs one unpatched vulnerability to flip control of a node. 
 (b)  The defender allocates defense resources to mitigate vulnerabilities by patching a component, thereby removing an attacker's exploitation possibility when attacked.  Note that only the defender knows the true patch state of each component.   
 }
 \label{fig:asymmetries}
\end{figure}

The cost of both types of actions were summarized in \S\ref{sec:CI}, and assumed to be commonly known. 

\subsection{Game Dynamics}
\label{sec:GD}
The game state at time $k$ is defined by the control states $\Gamma_{ik}$ for $j \in \{ 1 \ldots M \}$.  The available actions depend on the subset of $V$ reachable (i.e., $\{ j : r_{jk} = 1\}$, (which can further be dependent on games state and network topology).  

At time step $k$, the defensive actions are first used to update the $\phi$ variables as: 
\begin{equation}
\phi_{ijk} = \phi_{ij(k-1)} \otimes \zeta_{ijk}. 
\end{equation}
Note that the XOR operation will always flip the prior patch status\footnote{ Note that to weaken components by reverting a patch status is a {\it backdoor attack} and may be beneficial if it provides a means to re gain access a resource should an it be lost.}.

Next, at time step $k$, we will consider attacks. Recall that $\xi_{ijke}$ will indicate that vulnerability $i$ is attacked on machine $j$ at time $k$ by player $e$.  The attack will cost a resource $d_i$, and will be successful if $\phi_{ijk} = 0$, or otherwise futile if $\phi_{ijk} = 1$.
A successful attack will change the control state to that attacker,  starting at time step $k+1$.   A futile attack will not change any control state.  
Taking into account the actions of two players $e,e'$, we denote the full update to control states as:
\begin{equation}
\Gamma_{jke'} = \Gamma_{j(k-1)e'} \otimes  \phi_{ij(k-1)} \otimes \zeta_{ijke} \otimes \xi_{ijke'}.
\label{eq:GUF}
\end{equation}


\subsection{Agent Rationality and Conflicting Interest}
A scorebot rewards the player with a fixed reward $\mu$ for each unit of time a player controls the node. 
While players derive utility from unreachable nodes, reachable nodes at time step $k$ can potentially change control states, so player actions at time $k$ can reallocate the fixed rewards offered by reachable nodes.  Said differently, at time step $k$, players are reallocating control of ${\cal R}_k = \{ j : r_{jk} = 1 \} $, letting $R_k = | {\cal R}_k |$, players are reallocating $\mu R_k$ units of value at time step $k$.  
Players who control a node can perform defensive patch actions with the purpose to hold the node, while attackers will deploy exploits against ${\cal R}_k$ deemed most likely to change the nodes' control states.  

At time $k$, player interests over $\mu R_k$ units are completely opposed as a successful attack results in a direct transfer of utility from defender to attacker as in a zero sum quantity.

All player actions come at a cost, unlike \flipit where players can invest unlimited resources, we view constrained resource budgets for each time step $k$ as more realistic. 
To patch a vulnerability takes resources (time and cost), as does developing or deploying an exploit.

\subsection{ Utility Objectives and Resource Constraints } 
The utility of each player is an additive function aggregating time ownership of nodes.  
Each node will yield a contested resource at each time step for which its reachable. 
We make the more realistic assumption that utility is sought under a fixed and bounded resource budget (per times step).  Therefore, players plan action sequences to maximize the utility subject to the resource constraint.

\begin{equation}
U_e(T) = \sum_{k=0}^T \sum_{j = 1}^M \delta^k \mu_j \Gamma_{jke} \ \forall \ e \in {\cal A}    
\label{eq:U}
\end{equation}

Subject to the $k^{th}$ step resource limitation:

\begin{equation}
\sum_{i=1}^N\sum_{j=1}^M \left( d_i \xi_{ijk}^e + c_i \zeta_{ijk}^e \right) \leq C_e \ \forall \ k \in 1, \ldots T  \ \forall \ e \in {\cal A} 
\label{eq:C}
\end{equation}

Here $\mu_j$ generally models node value differentials. And $\delta \in (0, 1]$ is a standard future reward discount parameter used in Value Iteration algorithms.  

Rational agent will optimize their won individual objectives defined by equations~\ref{eq:U} subject to the system of constraints defined by equation~\ref{eq:C}, system dynamic update given by equations~\ref{eq:GUF}, will comprise our model an agent based system we term {\it dynamic network control game}.  

\subsection{Agent Based Model Summary} 
The novel dynamic network control game presented above resembles the informational uncertainties of \flipit, but refines the uncertainties to more realistic forms:  The attacker doesn't know component patch state and neither player will know the other's plan.  
We additionally avoid the unrealistic notion of unbounded budgets in \flipit, and return to an earlier notion of battling with constrained resources and bearing greater resemblance to the \colblotto game.  From this direction refinements include use of private information, and composition of components with a {\it weakest link logic} as it only takes one (of the many heterogeneous software components) successful exploit to flip the control state for any node. 
Our game posits agents as constrained optimizer of utility, and defines a logically complete set of update rules consistent with CTF games and adapted to the realism requirements mentioned before. 

\section{ Analysis and Methods } 
Analyzing play of the {\it dynamic network control game}, is a complex task due to its many parameters, however several computational paths forward are feasible.  First, numerical and simulation based approaches are able to test and evaluate a variety of agent strategies within specific environments, even those with complex parameters derived from real world networks with software inventory scanners and topological data.
Secondly, simplifying assumptions can be imposed to study specific cases and their resulting equilibrium.  Furthering this approach, a simulation and empirical game theory can assist to develop mathematical insights. 
Here, we take the second route and consider simplifying assumptions to deriving the \aknap problem and its special case for two players confined to attacker and defender actions respectively we term the \dknap.  
We start by reviewing the original knapsack problem. 

\subsection{Weighted Knapsack} 
Many constrained optimization problems over binary variables can be reduced to instance of a \wknap problem.  In \wknap, there exists a fixed set of $N$ objects having an associated weight $w_i$, and reward $\mu_i$.  An individual then selects objects to pack into a knapsack to maximize the net reward while not exceeding a fixed weight limit $W$.  
Letting $\theta_i$ be binary decision variables, the objective is the constrained optimization problem:
\begin{equation}
    \max \sum_{i = 1}^N \theta_i \mu_i \ \ : \ \  \sum_{i=1}^N \theta_i w_i \leq W 
    \label{eq:WKP}
\end{equation}
For uniform bound on the $W$, the \wknap problem can be solved efficiently for variables ${\bf \theta}$ using dynamic programming.  For increasing (unbounded) sequence in parameter $W$ the problem is known to reduce NP-complete problems, and has no general efficient solution unless $P=NP$.  
To address this, let ${\bf \theta}({\bf h}, {\bf m}, W )$ denote the solution to equation~\ref{eq:WKP}, and let ${\Tilde{\bf \theta}}({\bf h}, {\bf m}, W )$ denote the estimation of solution to equation (\ref{eq:WKP}) by feasible methods such as the dynamic programming.  
We have not encountering difficulty by using dynamic programming in the moderately sized network control problems our simulation software generates.  Thus, we observe that interesting network control problems seem to be solvable efficiently using binary programming, including the results shown for \dknap.


\subsection{Dueling Knapsacks}
A particularly relevant special case of the dynamic network control game, arising both in CTF as well as cyber operations, is when one player plays only as defense while the other plays only as attacker.   
Let the \dknap be the special case with two agents $a, d$  that act strictly in attack and defense mode respectively and play for a fixed time step $k$ (single round).  That is  $\xi_{ijkd} = 0 = \zeta_{ijka} \ \forall \ i, j, k  $.  
Note that for a single round, a rational defender should restrict actions to applying patches not removing them\footnote{Patch removal could be a viable option in CTF games where placing a backdoor into a contested nodes can ease acquisition should the node be compromised by another player. }. 

Letting, discount factor $\delta=1$, the first difference operator (in time) $\Delta_T$ applied to the utility equation~\ref{eq:U}, represents the utility gained by player $x$ for a fixed round starting at $T$ and ending at $T+1$ as: 
\begin{equation}
\Delta_T U_{x} = 
\sum_{j=1}^{M} \Gamma_{j(T+1)x}\mu_j = \sum_{j=1}^{M} \Gamma_{jTx} \left( \prod_{i}  \xi_i^a ( 1 - \zeta_i^d ) \right) \mu_j . 
\end{equation}
The first is by definition of first difference, the second equality is obtained by using the update function~\ref{eq:GUF}.
Next, dispensing with the time index, and noting that contended nodes are owned by the defender (i.e., $\Gamma_{jTd} = 1$, and letting $\zeta_i^a$ and $\xi_i^d$ denote more succinctly the actions of attacker $a$ and defender $d$, we derive:
\begin{equation}
\Delta U_{a} = \sum_{j=1}^{M} \left( \prod_{i}  \xi_i^a ( 1 - \zeta_i^d ) \right) \mu_j.
\label{eq:INU}
\end{equation}
From equation~\ref{eq:INU}, the utility gradient for attacker $a$ and defender $d$ is a zero sum (in complete conflict) for \dknap.  Said differently, any gain by attacker comes as a loss from defender. 
\begin{equation}
\Delta U_a = - \Delta U_{d}
\end{equation}

By noting that agents seek to optimize utility and by assuming the product found in equation~\ref{eq:INU} has a single factor, we can obtain the abstract \dknap problem.

\subsection{Abstract Dueling Knapsack}
For the \aknap problem, weights can generally be asymmetric or different for different players.  That is letting $w_i^a$ be the weight when object $i$ is used during attack, and $w_i^d$ is the weight when object $i$ is used during defense, its possible that $w^a_i \not = w^b_i$. 
The attacker $a$ seeks to exploit un-patched vulnerabilities, indicated by $\xi_i^{a} ( 1 -\zeta^d_i)$ . All exploits deployed are subject to the attacker resource constraint. 
\begin{equation}
\max \sum_{i=1}^N \left( \xi^{a}_i \left( 1 - \zeta_i^{d}\right) \right) \mu_i : \sum_{i=1}^N \xi^{a}_i w_i^a \leq W_{a}      
\label{eq:AK1}
\end{equation}
While the defender $e$ seeks to minimize the same sum by selecting patches subject to their resource constraint:
\begin{equation}
\min \sum_{i=1}^N \left( \xi^{a}_i \left( 1 - \zeta_i^{d}\right) \right) \mu_i : \sum_{i=1}^N \zeta^{d}_i  w_i^d \leq W_{d}      
\label{eq:AK2}
\end{equation}
Which is equivalent to:
\begin{equation}
\max \sum_{i=1}^N \left(  \zeta_i^{d} \xi^{a}_i \right) \mu_i : \sum_{i=1}^N \zeta^{d}_i w_i^d \leq W_{d}
\label{eq:AK3}
\end{equation}
   
This last form shows clearly the defender's need to be precise selecting to patch that which any attacker will next exploit\footnote{A multiplayer form of \aknap problem can be developed along similar lines, in equation~\ref{eq:AK1} the symbol $d$ can be replaced by $-a$ to indicated actions of any other player, likewise in equations~\ref{eq:AK2} and~\ref{eq:AK3} the symbol $a$ could be replaced by $-d$.  }.  

The \dknap problem can be constructed as either system of equations~\ref{eq:AK1} and~\ref{eq:AK2}, or system of equations~\ref{eq:AK1} and~\ref{eq:AK3}.  
Noting the assumption that each node is comprised of one component $i=j$, the constraints player roles in \dknap, and that $w_i^a, W_a$ correspond to $d_i, C_a$, while $w_i^d, W_d$ correspond to $c_i, C_d$, one can verify that equations~\ref{eq:U} and~\ref{eq:C} reduce to the above system of equations~\ref{eq:AK1} and~\ref{eq:AK3}.

We summarize the \dknap optimization problem in table~\ref{tab:t3}:  
\begin{table}[]
    \centering
    \begin{tabular}{|c|c|c|c|c|}
        \multicolumn{5}{c}{ \dknap (DK) Optimization } \\
        \hline 
         player   &  objective & mode & active resource constraint & solution \\
         \hline 
         \hline 
         $d$ &  $\min {\Bbb E}\left( \Delta U_{a} \right)$ &defender, $\xi_{ij}^d = 0$& $\sum_{i=0}^N w_i^d \zeta_{i}^d \leq W_d$ & patch plan: $\zeta^e$ \\
         \hline
        $a$ & $\max {\Bbb E} \left( \Delta U_{a} \right)$ &attacker $\zeta_{ij}^{a} = 0$& $\sum_{i=0}^N w_i^a \xi_{i}^{a} \leq W_a$ & exploit plan: $\xi^{a}$ \\
         \hline 
    \end{tabular}
    \caption{In \dknap the objectives for attacker and defender are completely opposed. The defender seeks to minimize expected gains while the attacker seeks to maximize them.  Each must select a plan within resource constraints. }
    \label{tab:t3}
\end{table}

\subsection{Approximate Solution of Dueling Knapsack by Fictitious Play}  
As a zero sum game, the existence of a mixed strategy equilibrium is assured, however for large $N$ its calculation can be shown to be prohibitive.  
Given $N$, there are up to $2^{N+1}$ feasible outcomes.  Both players, defender and attacker, can select actions as up to $2^N$ binary assignments (so long as they satisfy the weight constraints; i.e., equation~\ref{eq:C}).
The convex optimization problem required to calculate the equilibrium strategy is therefore lengthy to formalize, much less solve.
As such, each agent may consider reducing the search by exploring best response to a variety of plausible strategies the other may play.  

Rational players first goal is to optimize utility.  Due to the prohibitive computational task, players will heuristically search, so search becomes an important context for strategic reasoning, as players who can anticipate how the other player  search can maneuver.  We refer to this type of reasoning as secondary effects.  
To scrutinize a sequence of strategic maneuvers which players consider, we define an update procedure for \dknap and utilize it to calculate best response given a peer's strategy.

\subsubsection{Fictitious Play}
This type of reasoning involves fictitious play, and is performed by assuming the other's strategy, then updating equation~\ref{eq:AK1},~\ref{eq:AK2}, then optimizing the constrained optimization problem (eq~\ref{eq:U} and~\ref{eq:C}).
This method reduces the computational requirements to a computationally feasible instance of \wknap.  
Thus, while solving for the game equilibrium (minimax strategy) involves optimization over $2^{N+1}$ variables, fictitious play is a tool, and will calculate best response by fixing an opponent strategy and solving an instance of \wknap problem that reduces to a binary programming problem with $N$ variables.  We next show that this approach is computationally feasible and we detail the algorithms to calculate best responses for both attacker and defender.

Given an opponent's play (as a set of binary values), a player's {\it best response} is their choices (binary values)  that optimize their objective.  Let us observe how the knapsack problem changes, given an opponents strategy.  
For example:  given that the defender strategy ${\bm { \zeta }}$ which may have involved patching a fixed component $A$, (i.e., $\zeta_A^d = 1 $),  the attacker's best response ${\bm { \xi}}$ should avoid attacking $A$ and any other vulnerability that was patched, as it would be futile to attack component $A$, because its payoff is neutralized by the defender action.  This logic of avoidance can be interpreted from the term $(1 - \zeta_A^d)$, in equation~\ref{eq:AK1}, as it introduces a zero multiplier to $\mu_A$,  thus the attacker budget would be better spent on another component that remains un-patched.  Conversely, the defender seeks differently a logic of matching, as matching the actions of the attacker introduce zero multipliers to~\ref{eq:AK1}, or alternatively yield nonzero multipliers in~\ref{eq:AK3}.

In any case the logic of best response for \dknap has two phases: First to update the objective function treating other player strategies as data, next solve the \wknap (constrained optimization problem) by using dynamic programming.  Attacker's best response is calculated in algorithm~\ref{alg:a1}, defender's best response is calculated in algorithm~\ref{alg:a2}:

\begin{algorithm}
\caption{\sc Best-Response-A( ${\bf w, {\bm \mu} },W, {\bm \zeta}   $) }\label{alg:a1}
\begin{algorithmic}
\Require $\langle {\bf w}, {\bm \mu}, W , {\bm \zeta} \rangle $ \Comment{ Given costs/rewards, budget $W$, and defender strategy ${\bm \zeta }$}
\State ${\bf h } \gets ( {\bf 1 .- {\bm \zeta} } ).*{\bf \mu} $ \Comment{ phase i, update objective given defender choice }
\State ${\bf \xi} \gets {\bf \theta}( {\bf h}, {\bm w}, W )$ \Comment{ phase ii, solve Knapsack and return Best Response as ${\bm \xi}$}
\end{algorithmic}
\end{algorithm}

\begin{algorithm}
\caption{\sc Best-Response-D( ${\bf w, {\bm \mu} },W, {\bm \xi}   $) }\label{alg:a2}
\begin{algorithmic}
\Require $\langle {\bf w}, {\bm \mu}, W , {\bm \zeta} \rangle $ \Comment{ Given cost/rewards, budget $W$, attacker strategy ${\bm \xi }$}
\State ${\bf h } \gets ( {\bm \xi} ).*{\bf \mu} $ \Comment{ phase i, update objective given attacker choice}
\State ${\bf \zeta} \gets {\bf \theta}( {\bf h}, {\bm w}, W )$ \Comment{ phase ii, solve Knapsack and return Best Response as ${\bm \zeta}$}
\end{algorithmic}
\end{algorithm}

\subsubsection{Extending Fictitious Play to estimate game value.}
With the tools of best response, i.e., algorithms~\ref{alg:a1} and~\ref{alg:a2}, effective strategies can be generated as a type of virtual pre-play in preparation to the game.  
Noting that another property of Zero Sum games is that at equilibrium (minimax) the dual objectives match in value to the {\it game value}.  
Thus, players who emulate the actions of their opponent can derive outcome scores for both players to estimate the game value.  Therefore, Fictitious play can in some cases identify robust strategies or that certain strategies are good enough (for example, a strategy whose safety level is within 8\% of the game value, could be considered good enough to justify looking no further).  We provide an example sequence of fictitious play.

\subsection{Secondary Effects for Rational Agents in Dueling Knapsack }

Strategic agents form strategies not only over problem parameters, but also with respect to the strategies that other agents are likely to play.  These secondary effects involve optimization and maneuvering to the likely strategic play of other agents.  Here we show that they can determine a sequence of best response strategic forms, and conclude with optimization of expected utility given a belief probability of other player's frequency of patch/exploit actions. 

During pre-play or planning, we illustrate the type of reasoning that fictitious play facilitates.  
At step $0$, each agent could start with a simple random (i.e., {\it oblivious}) strategy; that is, defense randomly select a set of patches to apply until the resource constraint is exhausted, and likewise offense likewise randomly selects a set of exploits until its resource constraint is exhausted.  
Let $\zeta_0, \xi_0$ represent these simple random selection processes, which select components equally in probability. 

As step $1$, both agents can reason that random strategies are a natural starting strategy for the other, and so, each agent may seek improvements, by optimization of one's own resource use under the assumption that the other randomly selects patches/exploits.  For attacker, this means using common knowledge ${\bf w}, {\bm \mu}$, their own budget $W_a$ and setting $\zeta = [1, 1, \hdots, 1]$, then calculating  
$$\xi_1 = \text{{\sc Best-Response-A}}( {\bf w}, {\bm \mu}, W_a, [1,1,\hdots, 1] ).$$ 
Likewise for defender, a budget constrained hazard minimizing strategy uses common knowledge ${\bf w}, {\bm \mu}$, defense budget $W_d$ and setting $\xi = [1, 1, \hdots, 1 ]$, then calculates
$$\zeta_1 = \text{{\sc Best-Response-D}}( {\bf w}, {\bm \mu}, W_a, [1,1,\hdots, 1] ).$$

Noting that the common knowledge also enables each player to solve the other's optimization problem, we further reason that players may maneuver in the $k^{th}$ step.  
The attacker will have calculated $\xi_{k-1}$, but in an attempt to stay a step ahead of the defender, will also calculate $\zeta_{k-1}$ to see if maneuvering can result in a higher score with: 
$$\xi_k = \text{{\sc Best-Response-A}}( {\bf w}, {\bm \mu}, W_a, \zeta_{k-1} ),$$ 
Likewise, the defender will have calculated $\zeta_{k-1}$, but in an attempt to stay a step ahead of the attacker, will also calculate $\xi_{k-1}$ to see if maneuvering can result in a higher score with:
$$\zeta_k = \text{{\sc Best-Response-A}}( {\bf w}, {\bm \mu}, W_a, \xi_{k-1} ),$$ 

Essentially, both players using computational have access to a sequence of strategies that can be derived as best response during pre-play planning.  While at first these sequences may be constructed by players in attempts to seek maneuvers to gain advantage.  
Introspective players will realize that other rational introspective players will derive the same sequence.  Let the {\it fictitious play path} be the derived sequence:  
\begin{equation}
( \zeta_0, \xi_0 ), ( \zeta_1, \xi_1 ), ( \zeta_2, \xi_2) , \hdots 
\label{eq:fpseq}
\end{equation}
 and note that it can be considered a commonly known search path for the prohibitive search for minimax strategy.

As such, the sequence may be understood as a measurement tool to mediate outcomes in an otherwise prohibitive optimization task.  Analysis of the game may next turn to inference of game value, variation afforded to maneuvering, sequence cycle length, attainment of game value within the sequence, and the rates at which player's can close in on game value by search.  We illustrate answers to these questions in the results section to follow.

In another type of analysis where expected rewards are optimized over belief probabilities, the sequence in~\ref{eq:fpseq} can play an important roll, as it can be used to estimate probability that players will patch/exploit any particular option by using empirical frequencies.  
That is for attacker, let ${p_A}$ be:
\begin{equation}
{p_A} = \lim_{ n \rightarrow \infty } \frac{ \sum_k \zeta_{kA} }{n},
\label{eq:pae}
\end{equation}
where $\zeta_{kA}$ is the binary variable determining if defender patches $A$ for strategy $\zeta_k$.  
For defender, let ${q_A}$ be: 
\begin{equation}
{q_A} = \lim_{ n \rightarrow \infty } \frac{ \sum_k \xi_{kA} }{n},
\label{eq:qae}
\end{equation}

We present the general expected optimization, given a belief probability over the opponent's patch/exploit frequencies as algorithms~\ref{alg:a3} and~\ref{alg:a4}: 

\begin{algorithm}
\caption{\sc Expected-Optimal-A( ${\bf w, {\bm \mu} },W, {\bf p}   $) }\label{alg:a3}
\begin{algorithmic}
\Require $\langle {\bf w}, {\bm \mu}, W , {\bf p} \rangle $ \Comment{ Given defender probability ${\bf p }$ for patches.}
\State ${\bf h } \gets ( {\bf 1 .- {\bf p} } ).*{\bf \mu} $ \Comment{ phase i, update objective represents expected rewards given ${\bf p}$ }
\State ${\bf \xi} \gets {\bf \theta}( {\bf h}, {\bm w}, W )$ \Comment{ phase ii, solve Knapsack and return Expected Optimal as ${\bm \xi}$}
\end{algorithmic}
\end{algorithm}

\begin{algorithm}
\caption{\sc Expected-Optimal-D( ${\bf w, {\bm \mu} },W, {\bf q}   $) }\label{alg:a4}
\begin{algorithmic}
\Require $\langle {\bf w}, {\bm \mu}, W , {\bf q} \rangle $ \Comment{ Given attacker probability ${\bf q }$ for port of attack}
\State ${\bf h } \gets ( {\bf q} ).*{\bf \mu} $ \Comment{ phase i, update objective represents expected reward given ${\bf q}$}
\State ${\bf \zeta} \gets {\bf \theta}( {\bf h}, {\bm w}, W )$ \Comment{ phase ii, solve Knapsack and return Best Response as ${\bm \zeta}$}
\end{algorithmic}
\end{algorithm}

\section{Results}
Our study of \dknap concludes with insights from a process of mathematical simulation.  We discover two insights: first, that maneuvering (considered in fictitious play) will cycle rather quickly.  Second, we find that players can use fictitious play to estimate game value and determine reasonable strategies. Further, expectation maximization should tighten the bounds for game value estimation even more.  Taken together, these insights suggest that while secondary effects may offer slight maneuverability for players, they ultimately are not game changing.

\subsection{Maneuvering and Estimating Game Value in Dueling Knapsack}
To understand the special case of \dknap, we design simulation software written in the Julia programming language to generate random networks composed of nodes comprised of vulnerable components.  
We use statistical distributions to generate: vulnerabilities within components, random cost structure for patching, and random cost structure for exploitation.   
To solve knapsack instances, we use binary programming routines provided by JuMP~\cite{DunningHuchetteLubin2017}.  We highlight several interesting deeper mathematical questions and summarize the insights from simulating realistic size problems.

For the following discussion we generate networks of 100 machines, 70 vulnerabilities, Each node has a number of distinct vulnerabilities ranging from zero to 10.  The vulnerabilities are assigned hazard scores ranging uniformly between 6 and 100.  Patch cost distribution ranges uniformly between 100, and exploit development ranges uniformly between 100 and 700.  And each player has a fixed budget constraint of 800.  We have not yet performed a wide statistical sampling of all such configurations, however,  we have generated dozens of problem instances to observe the following characteristics.

\subsubsection{Limited Maneuvering} 
\begin{figure}[htb]
\centering
 \subfigure[cycling strategic sequence]{\includegraphics[height=34mm]{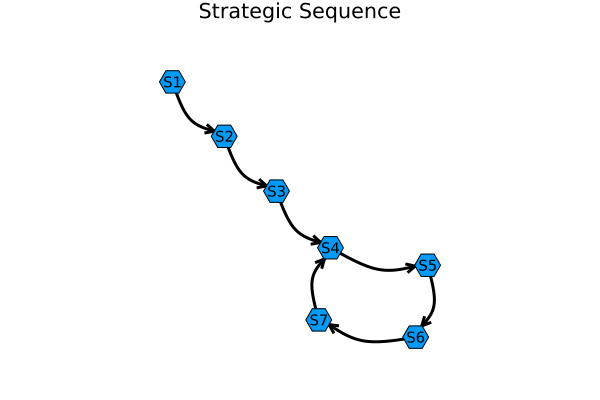} \label{fig:sol1}}
\subfigure[Payoff matrix]{\includegraphics[height=34mm]{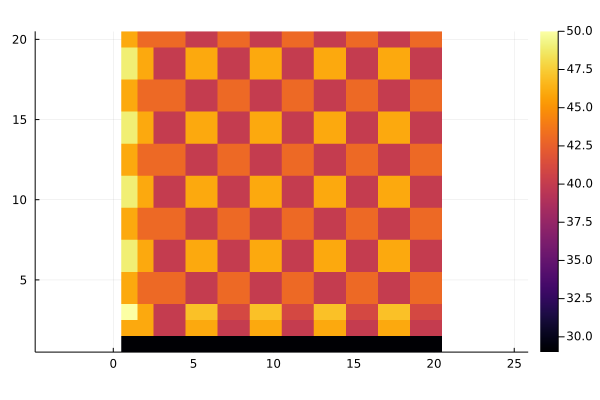}\label{fig:sol2}}  
\caption{Bounding the Secondary effects and how agents may utilize common knowledge in fictitious play.  We generated networks with 100 machines, 70 vulnerabilities, and calculate the fictitious play sequence starting from random strategies.  Generally, as in the image above, we observe that fictitious play cycles within a few steps.  In (a) we illustrate an example strategy sequence cycling, each node represents a distinct tuple of strategies identified by binary variable selection for $2^{N+1}$ variables.  
In (b) we collect and visualize the payoff outcomes of all pairs $7 \text{Chose} 2$ strategies discovered in part (a).  The associated zero-sum payoff matrix can provide both players a means to estimate game value. }
\end{figure}

Using our simulation approach, we find that the fictitious play sequence (\ref{eq:fpseq}) cycles rather quickly (see image~\ref{fig:sol1}) even in large problems, thereby suggesting limits to what a player can accomplish with maneuvering in \dknap.   
Theoretically, a cycle in assured due to the finite and discrete nature of knapsack problems established by player selections of binary variables.  While we can bound cycle length to $2^{N+1}$, the total number of actions available in the game, simulation suggests sequence~\ref{eq:fpseq} will cycle and find limit cycle in only a few steps.  

The cycle length and number of steps needed to enter the cycle may have a statistical relationship to the distribution and parameters of networks.  We leave these broader questions open for future work, but comment here on some intuitive aspects of networks which can help form insights as to why cycles are small and quick to come by.

First, constraints make some actions infeasible, interesting problems will engage the constraints, thus ruling out infeasible actions because all strategies in the fictitious play sequence are required to be feasible.  Thus, the constraints reduce the bound on cycle length.   
Secondly, interesting problems will also feature variation in costs, and hazards, and variation will induce priorities for both utility seeking players.  Said differently, variations in parameters assure variation in utility and sparsity of the best responses, therefore rational players will never elect to put poor performing strategic options into a fictitious sequence, thereby reducing cycle length dramatically.  Still interesting mathematical questions arise, such as how parameters of networks can relate to sparsity of the best responses.

\subsubsection{Estimation of Game Value}
\begin{figure}[htb]
\centering
 \includegraphics[height=60mm]{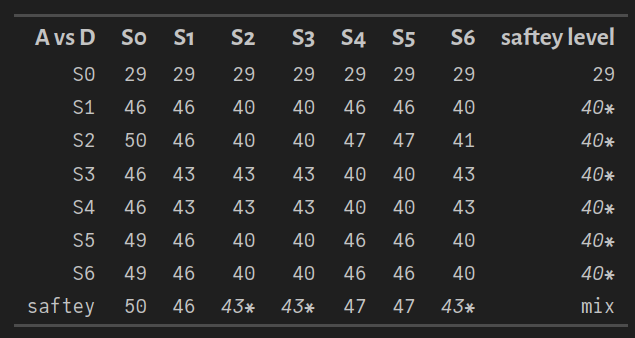}\label{fig:maximin}
 \caption{{\it Strategic Stabilization at Maximin:} Zero sum games have always have mixed strategy equilibrium, but may also have pure equilibrium's when maximin and minimax are equal.  In this example, where analysis is restricted to strategies discovered with fictitious play the attacker has a maximin strategy to guarantee a gain of $40$ nodes (or more), the defender can guarantee a loss of no more than 43.  These values bound the game value and can be used to evaluate and optimize network designs for cyber security.  These values represent relative difference of less than 7.5\% of game value. }
 \label{fig:pwnitgt}
\end{figure}

Zero-sum games admit to a mixed strategy equilibrium where the attacker minimax strategy (optimizer of equation~\ref{eq:AK1}) value equals that of the defender maximin strategy (optimizer of equation~\ref{eq:AK2}) value, the matching value is termed the game value at equilibrium.   While exploration of all strategies remains computationally infeasible, the fictitious play sequence can be used for exploratory search, and the payoffs can be evaluated using all pairs of strategies discovered along the fictitious play path (see image~\ref{fig:sol2}). Given the need for sampling strategies, players could generate simulated play among sampled strategies and perform standard analysis of the game based on the limited sample.  
Standard Zero-Sum game analysis such as calculation of a strategy's {\it saftey level}, defined to be minimum score (for attacker) that an attacker strategy can ensure, and the maxiumum score (for attacker) that a defender strategy can yield, could be applied to the sampled strategies, done in figure ~\ref{fig:pwnitgt}.   Interestingly, both sides can determine strategies within $7.5\%$ of game value, arguably good enough given the computational infeasibility of exhaustive search.     

One additional method which players can use to tighten the estimate bounds for game value would entail calculating frequencies in which each component is patched/exploited in fictions play by using equations~\ref{eq:pae} and ~\ref{eq:qae}, convergence is assured by our earlier observation that fictitious play sequence (equation~\ref{eq:fpseq}) cycles.  Moreover, small cycle length further expediates convergence rate.  Using these estimated frequency values, algorithms~\ref{alg:a3} and~\ref{alg:a4} can be used by attacker and defender respectively to calculate mixed strategies.

\subsection{ Dueling Knapsack Application to Cyber Operations } 
The \dknap problem is closely related to the momentary considerations during cyber operations and CTF competitions.  Generally, it models the tasks of offense and defense are divided into separate concerns and captures the usual case of defense of a fixed network while offense attempts to compromise any reachable nodes.  Reachability and component enumerations can be performed by scanning tools that conduct node and software component inventory.   Vulnerability inventory assessments and patch management also have automating tools, as well as automated exploit generation and penetration testing to estimate problem parameters.  Weights and costs increasingly is becoming common information Common Vulnerability Exposure (CVE) and other metric sources can be used to form commonly known hazard scores $\mu$ and cost functions ${\bf w^a, w^d}$.  

\subsection{Simulation and Decision Support Software}
While \dknap provides intuition for a single stage in these games, the dynamic game form is far more complex.  
Like in the \dknap problem, the role of fictitious play and simulation offer interesting possibilities for the complex dynamic agent based game model presented herein.  Modeling networks is not difficult as their formal descriptions using graphs, hazards and costs can generate Agent Based System Models, and the \aknap game of equation~\ref{eq:U} and~\ref{eq:C}.  Therefore, we see a role for software that  considers fictitious play and simulation to form tools to explore strategic actions in cyber operational scenarios.  Below in figure~\ref{fig:sim} we show how software tools which consider optimization strategy more generally could work.  

\begin{figure}[htb]
\centering
 \subfigure[Initial game state]{\includegraphics[height=40mm]{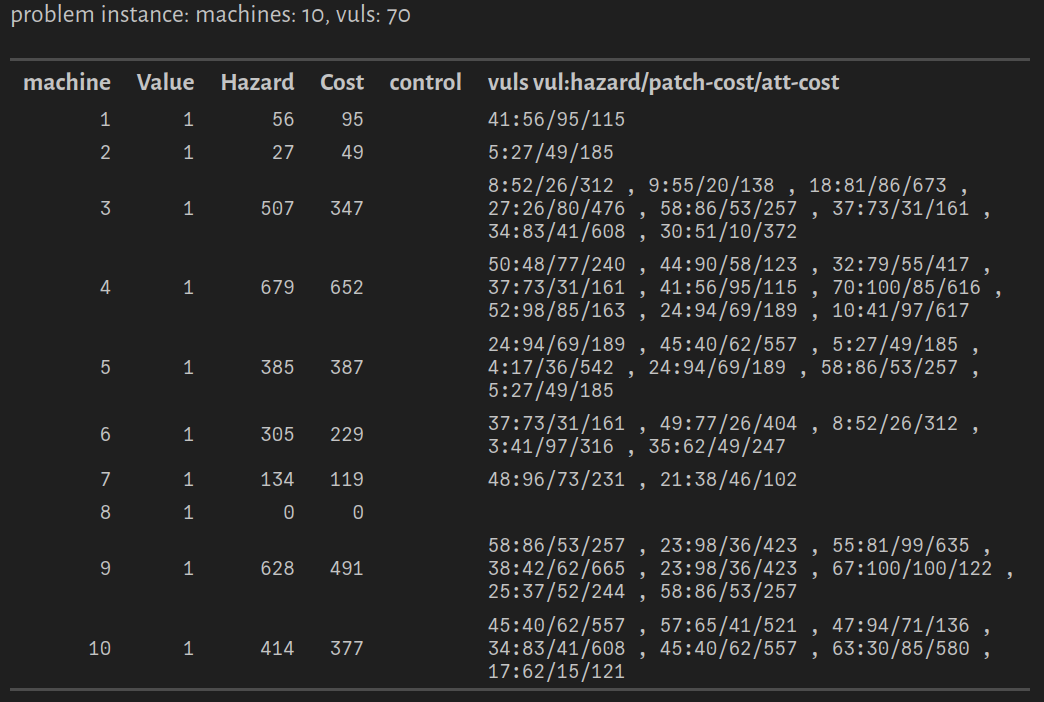}\label{fig:pi1}}
\subfigure[One time step later]{\includegraphics[height=40mm]{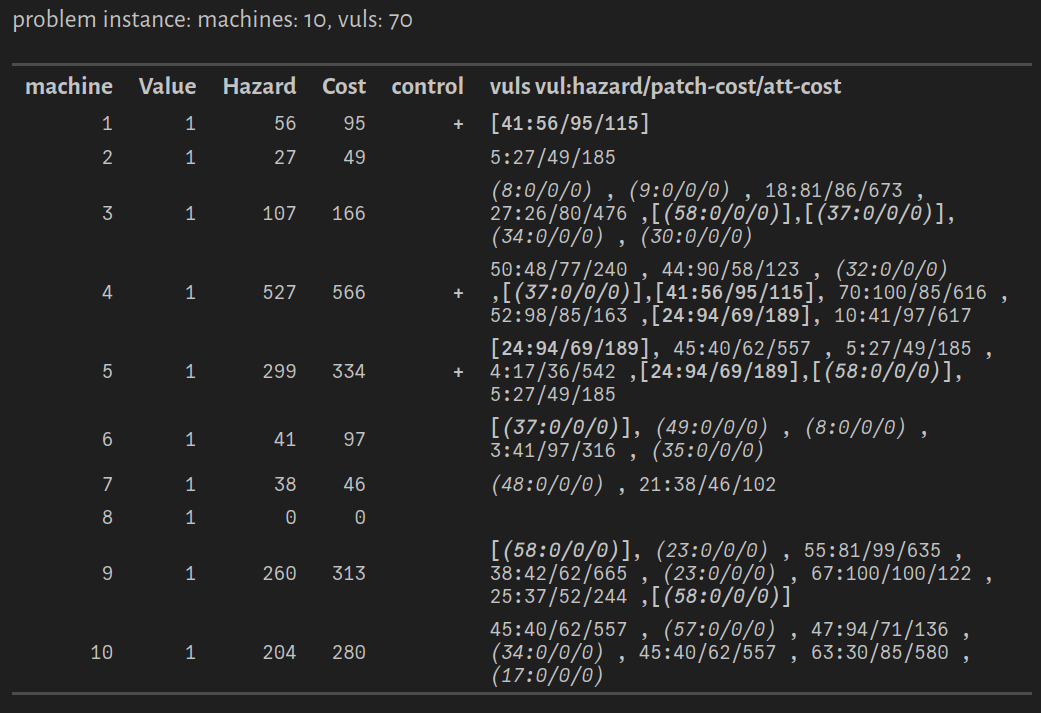}\label{fig:pi2}}  
\caption{Encoding the networks into the game model can be automated, In (a) a random initial game state is provided, it enumerating machines and vulnerabilities along with associated hazard and costs for exploit/patch, in (b) we observe the effects of player actions: '+' indicates flipped control states, '($x$)' indicates patch actions on $x$, '[$y$]' indicates exploit actions attempt on $y$.  Notice that game states are updated to enable simulation and evaluation of multi-step strategies. }
\label{fig:sim}
\end{figure}

\subsection{Summary} 
We envision several applications of decision support software employing simulation of \aknap. 
First, as a measurement tool, simulation can be used to evaluate game outcomes in a variety of ways, including for a specified network subset and software component configuration.  As such, it provides a design and assessment tool which allows a network engineer to consider the vulnerability surface at each potential stage of a network attack.  Applying \aknap can determine effects of different network typologies, 
or evaluate security hardened postures, such as minimizing or containing services.

Second, this methodology can be applied in dynamic scenarios such as cyber operations or within CTF game competitions.  In this capacity, the methodology enables fictitious play where operators could potentially play out several moves into the future to evaluate simulated possibilities, or possibly even to receive decision assistance.  One aspect of this type of application and a direction we plan to explore in the near future work will involve educational support tools which may help cyber operators learn more effective use of resources.  

\subsection{Simulation Code} 
We attach our code in the appended materials. 
Our program has been run on very large simulated networks with thousands of nodes and vulnerabilities.  
Little prevents the code from working on actual network data when encoded using standard data structures of graphs and maps.  
In our simulations the vulnerability scores, costs $w^d, w^a$ are generated using statistical distributions. 
As future work, we plan to apply this program to the use case of vulnerability scanners such as Nessus~\cite{nessus} and OpenVAS vulnerability scanner~\cite{openvas}. Additionally, we assert the use of CVSS scores~\cite{cvss} will increase the accuracy of these values.

\section{Conclusion}
Stemming from research showing how combinatorial optimizations (such as solving instances of \wknap) can result in better patch planning, we set out to extend the same type of reasoning to the attacker objectives.
Commonly known to all agents are hazard scores associated with vulnerabilities, as well as their patch (remediation) and exploit costs.  This information provides important and realistic information about the risks that certain vulnerabilities have within systems, and would naturally be leveraged by all agents in planning defense and attack. 
Common knowledge and common awareness of the same raises a simple but basic question of secondary reasoning:
{\it if each player is capable of solving the other's knapsack problem, might they attempt to out maneuver the other's solution?}  
We resolve this question by formalizing agent based models for network control, which feature bounded resources such as \colblotto and information asymmetric aspects such as \flipit.  
Then, we capture the single time step game as instances of a novel distributed optimization problem we term \aknap and the special case of \dknap.  
To evaluate the secondary effects at play in \dknap,
we develop an analysis tool that calculates best-response reducing it to instances of knapsack.  Using this tool, we can expand the chain of secondary reasoning, collecting strategies that could be considered in fictitious play.  We determine that the secondary reasoning offers limited maneuvering, as best responses in fictitious play eventually cycle and loop back into previously explored strategies.  Cycles are explored and found to occur rather quickly, thereby bounding the strategic options but also offering a means to estimate game values and the possibility of identifying reasonable strategies. 
 
The model and methodology can be applied to design problems as well as decision support systems, however that remains as future work.  
Perhaps another interesting question that the model suggests is the use of misinformation.  Given that individual capabilities and costs structures naturally vary, a question becomes how might errors or even manipulated common knowledge affect \aknap and what are the implications for the single time step game let alone the dynamic one.  
Another means to asking a related question, is by adding noise and distortion to game data as well as considering agent intentional noise such as trembling hand strategies. 
Additionally, many interesting mathematical questions for \aknap equilibria arise and remain open:  What are the statistics of cycle length, and how do they depend on distributions (of hazards, costs of patching and exploiting)?  How do they depend on the distribution of vulnerabilities within networks or even network topological aspects?  
These questions may seem to offer limited resolution in small networks, but how might they scale in larger networks?  

While we focused on the impact of common information such as CVE scores, this approach could be extended to consider situations where one player has non-public intelligence about the opponent to solve their knapsack problem. Research in this direction could help calculate the value of such intelligence. Conversely, an expanded model could quantify the defensive value of non-public information, such as being able to patch a novel vulnerability that is not yet known to the opponent.


%
%
%
%

\bibliographystyle{plain}
\bibliography{main}

\begin{thebibliography}{10}

\bibitem{cvss}
{Common Vulnerability Scoreing System SIG}.
\newblock \url{https://www.first.org/cvss/}.
\newblock Accessed: 2023-11-29.

\bibitem{nessus}
{Nessus Vulnerabilty Scanner}.
\newblock \url{https://www.tenable.com/products/nessus}.
\newblock Accessed: 2023-11-29.

\bibitem{openvas}
{Open Vulnerability Assessment Scanner}.
\newblock \url{https://www.openvas.org/}.
\newblock Accessed: 2023-11-29.

\bibitem{ctf}
Capture The~Flag 101.
\newblock {CTF 101}.
\newblock \url{https://www.ctf101.org }, 2022.

\bibitem{beier2004random}
Ren{\'e} Beier and Berthold V{\"o}cking.
\newblock Random knapsack in expected polynomial time.
\newblock {\em Journal of Computer and System Sciences}, 69(3):306--329, 2004.

\bibitem{bock2018king}
Kevin Bock, George Hughey, and Dave Levin.
\newblock King of the hill: A novel cybersecurity competition for teaching
  penetration testing.
\newblock In {\em 2018 USENIX Workshop on Advances in Security Education (ASE
  18)}, 2018.

\bibitem{casey_epi2014}
William Casey, Rhiannon Weaver, Leigh Metcalf, Jose~Andre Morales, Evan Wright,
  and Bud Mishra.
\newblock Cyber security via minority games with epistatic signaling.
\newblock In {\em Proceedings of the 8th international conference on
  bioinspired information and communications technologies}, pages 133--140,
  2014.

\bibitem{cavusoglu2006economics}
Huseyin Cavusoglu, Hasan Cavusoglu, and Jun Zhang.
\newblock Economics of security patch management.
\newblock In {\em Workshop on Economics of Information Security (WEIS 2006)}.
  Citeseer, 2006.

\bibitem{DunningHuchetteLubin2017}
Iain Dunning, Joey Huchette, and Miles Lubin.
\newblock Jump: A modeling language for mathematical optimization.
\newblock {\em SIAM Review}, 59(2):295--320, 2017.

\bibitem{etesami2019dynamic}
S~Rasoul Etesami and Tamer Ba{\c{s}}ar.
\newblock Dynamic games in cyber-physical security: An overview.
\newblock {\em Dynamic Games and Applications}, 9(4):884--913, 2019.

\bibitem{ferdowsi2017colonel}
Aidin Ferdowsi, Walid Saad, Behrouz Maham, and Narayan~B Mandayam.
\newblock A colonel blotto game for interdependence-aware cyber-physical
  systems security in smart cities.
\newblock In {\em Proceedings of the 2nd international workshop on science of
  smart city operations and platforms engineering}, pages 7--12, 2017.

\bibitem{cvssinattackgraphs}
Laurent Gallon and Jean-Jacques Bascou.
\newblock {Using CVSS in attack graphs}.
\newblock pages 59--66, 2011.

\bibitem{goohs2022reducing}
Jonathan Goohs, Ray Mier, Paul Deist, and William Casey.
\newblock Reducing attack surface by learning adversarial bag of tricks.
\newblock In {\em Workshop on Economics of Information Security (WEIS 2022)},
  2022.

\bibitem{greige2020reinforcement}
Laura Greige and Peter Chin.
\newblock Reinforcement learning in flipit.
\newblock {\em arXiv preprint arXiv:2002.12909}, 2020.

\bibitem{gross1950continuous}
Oliver Gross and Robert Wagner.
\newblock A continuous colonel blotto game.
\newblock Technical report, Rand Project Air Force Santa Monica Ca, 1950.

\bibitem{gupta2014three}
Abhishek Gupta, Galina Schwartz, C{\'e}dric Langbort, S~Shankar Sastry, and
  Tamer Ba{\v{r}}ar.
\newblock A three-stage colonel blotto game with applications to cyberphysical
  security.
\newblock In {\em 2014 American Control Conference}, pages 3820--3825. IEEE,
  2014.

\bibitem{htb}
{Hack The Box}.
\newblock {Hack The Box: Hacking Training For the Best}.
\newblock \url{https://www.hackthebox.com}, 2022.

\bibitem{jacobs2023enhancing}
Jay Jacobs, Sasha Romanosky, Octavian Suciu, Ben Edwards, and Armin Sarabi.
\newblock Enhancing vulnerability prioritization: Data-driven exploit
  predictions with community-driven insights.
\newblock In {\em 2023 IEEE European Symposium on Security and Privacy
  Workshops (EuroS\&PW)}, pages 194--206. IEEE, 2023.

\bibitem{labib2015colonel}
Mina Labib, Sean Ha, Walid Saad, and Jeffrey~H Reed.
\newblock A colonel blotto game for anti-jamming in the internet of things.
\newblock In {\em 2015 IEEE global communications conference (GLOBECOM)}, pages
  1--6. IEEE, 2015.

\bibitem{laszka2014flipthem}
Aron Laszka, Gabor Horvath, Mark Felegyhazi, and Levente Butty{\'a}n.
\newblock Flipthem: Modeling targeted attacks with flipit for multiple
  resources.
\newblock In {\em International Conference on Decision and Game Theory for
  Security}, pages 175--194. Springer, 2014.

\bibitem{liu2021flipit}
Zhaoxi Liu and Lingfeng Wang.
\newblock Flipit game model-based defense strategy against cyberattacks on
  scada systems considering insider assistance.
\newblock {\em IEEE Transactions on Information Forensics and Security},
  16:2791--2804, 2021.

\bibitem{mcmorrow2010science}
Dale McMorrow.
\newblock Science of cyber-security.
\newblock Technical report, The MITRE Corporation, 2010.

\bibitem{merkle1978hiding}
Ralph Merkle and Martin Hellman.
\newblock Hiding information and signatures in trapdoor knapsacks.
\newblock {\em IEEE transactions on Information Theory}, 24(5):525--530, 1978.

\bibitem{min2017defense}
Minghui Min, Liang Xiao, Caixia Xie, Mohammad Hajimirsadeghi, and Narayan~B
  Mandayam.
\newblock Defense against advanced persistent threats: A colonel blotto game
  approach.
\newblock In {\em 2017 IEEE international conference on communications (ICC)},
  pages 1--6. IEEE, 2017.

\bibitem{munaiah2016vulnerability}
Nuthan Munaiah and Andrew Meneely.
\newblock Vulnerability severity scoring and bounties: Why the disconnect?
\newblock In {\em Proceedings of the 2nd International Workshop on Software
  Analytics}, pages 8--14, 2016.

\bibitem{oakley2019mathsf}
Lisa Oakley and Alina Oprea.
\newblock Qflip: An adaptive reinforcement learning strategy for the flipit
  security game.
\newblock In {\em International Conference on Decision and Game Theory for
  Security}, pages 364--384. Springer, 2019.

\bibitem{savin2023battle}
Georgel~M Savin, Ammar Asseri, Josiah Dykstra, Jonathan Goohs, Anthony
  Melaragno, and William Casey.
\newblock Battle ground: Data collection and labeling of ctf games to
  understand human cyber operators.
\newblock In {\em Proceedings of the 16th Cyber Security Experimentation and
  Test Workshop}, pages 32--40, 2023.

\bibitem{schwartz2014heterogeneous}
Galina Schwartz, Patrick Loiseau, and Shankar~S Sastry.
\newblock The heterogeneous colonel blotto game.
\newblock In {\em 2014 7th international conference on NETwork Games, COntrol
  and OPtimization (NetGCoop)}, pages 232--238. IEEE, 2014.

\bibitem{shamir1982polynomial}
Adi Shamir.
\newblock A polynomial time algorithm for breaking the basic merkle-hellman
  cryptosystem.
\newblock In {\em 23rd Annual Symposium on Foundations of Computer Science
  (SFCS 1982)}, pages 145--152. IEEE, 1982.

\bibitem{tirenin1999concept}
Walt Tirenin and Don Faatz.
\newblock A concept for strategic cyber defense.
\newblock In {\em MILCOM 1999. IEEE Military Communications. Conference
  Proceedings (Cat. No. 99CH36341)}, volume~1, pages 458--463. IEEE, 1999.

\bibitem{van2013flipit}
Marten Van~Dijk, Ari Juels, Alina Oprea, and Ronald~L Rivest.
\newblock Flipit: The game of “stealthy takeover”.
\newblock {\em Journal of Cryptology}, 26(4):655--713, 2013.

\bibitem{wang2016recent}
Dong Wang, Zidong Wang, Bo~Shen, Fuad~E Alsaadi, and Tasawar Hayat.
\newblock Recent advances on filtering and control for cyber-physical systems
  under security and resource constraints.
\newblock {\em Journal of the Franklin Institute}, 353(11):2451--2466, 2016.

\bibitem{zhang2014stealthy}
Ming Zhang, Zizhan Zheng, and Ness~B Shroff.
\newblock Stealthy attacks and observable defenses: A game theoretic model
  under strict resource constraints.
\newblock In {\em 2014 IEEE Global Conference on Signal and Information
  Processing (GlobalSIP)}, pages 813--817. IEEE, 2014.

\end{thebibliography}


\end{document}